\documentclass{article}
\usepackage{times}
\usepackage{amsfonts}
\usepackage{graphicx}
\begin{document}
\noindent
{\Large   ANALYSIS OF OBSERVABLES IN FLRW SPACETIME}
\noindent

\vskip.5cm
\noindent
{\bf P. Fern\'andez de C\'ordoba}$^{1}$, 
{\bf S. Gavasso}$^{2}$ and {\bf  J.M. Isidro}$^{3}$\\
Instituto Universitario de Matem\'atica Pura y Aplicada,\\ Universitat Polit\`ecnica de Val\`encia, Valencia 46022, Spain\\
$^{1}${\tt pfernandez@mat.upv.es}, $^{2}${\tt sgavas@alumni.upv.es},\\
$^{3}${\tt joissan@mat.upv.es}\\

%\vskip.5cm
%\noindent
%\today
\vskip.5cm
\noindent
{\bf Abstract}  We consider a positively curved FLRW spacetime as a background in which a nonrelativistic quantum particle propagates according to the Schroedinger equation. The probability fluid for the corresponding quantum states is taken as a model for the cosmological fluid filling this FLRW Universe. The Hamiltonian operator governing this fictitious particle is proportional to the Laplacian operator derived from the FLRW metric, while the mass of the particle equals the overall matter of the Universe (baryonic and dark). A complete, orthonormal set of quantum eigenstates of the Hamiltonian is obtained. Restricting to radially symmetric states, the latter are then used to compute matrix elements and expectation values of two observables for which quantum operators are identified, namely, the cosmological constant and the gravitational Boltzmann entropy. This entropy is regarded as corresponding to a positively--curved FLRW geometry when the cosmological fluid filling the Universe occupies a given quantum state.

%\tableofcontents

\section{Introduction}\label{entro}

Over the years it has become increasingly patent that gravity is amenable to a thermodynamic description \cite{JACOBSON, PADDY1}. The possibility of describing gravity in thermodynamic terms has led to the suggestion that spacetime must be a derived concept, rather than a primary one. Numerous proposals concerning the emergent nature of spacetime have been put forward \cite{FISCHER, STAT, PADDY2, SINGH}; for extensive literature see, {\it e.g.}\/, ref. \cite{MOUSTOS}. Instead of viewing Einstein's equations as a geometric description of gravity, the thermodynamics of spacetime proposes that they be understood as the equilibrium condition of a thermodynamic system, where horizons act as the fundamental thermodynamic components. This leads to the profound conclusion that gravity itself must be a macroscopic, emergent phenomenon arising from the microscopic, statistical behavior of the fabric of spacetime.

In particular, the notion of entropy appears to be of central importance to gravity \cite{PENROSE}; for sample references see, {\it e.g.}\/, \cite{PERRY, ELLIS}. The conceptual as well as the observational aspects of the entropy of the Universe have long been the subject of keen attention \cite{ASTROPH, FRAMPTON, CENSUS}. In this context one must mention entropic gravity as especially suited to analyse spacetime from a  thermodynamic point of view \cite{VERLINDE1, VERLINDE2}. It almost follows that one must apply entropic techniques to problems in cosmology, too \cite{UKRAINE, ODINTSOV1, ODINTSOV2}.

Our present knowledge describes the Universe fairly accurately by means of a Friedmann-Lema\^{\i}tre-Robertson-Walker (FLRW) geometry \cite{WEINBERG2}, with a tiny but positive cosmological constant $\Lambda$ driving its accelerated expansion \cite{PERLMUTTER, RIESS}.  At the same time, while spacetime requires general relativity for its precise comprehension, in many situations the matter contents within it (baryonic plus dark) appears to be reasonably well described in Newtonian terms, {\it i.e.}\/, in terms of nonrelativistic dynamics.

In this letter we adopt the viewpoint just described and regard the cosmological fluid as a Newtonian fluid within a positively--curved FLRW geometry. In turn, this latter fluid is described quantum--mechanically {\`a} la Madelung, {\it i.e.}\/, as a quantum probability fluid. This feature allows one to define two quantum--mechanical observables that are relevant for a thermodynamic analysis of the Universe, namely, the cosmological constant operator and the entropy operator. In this way we extend previous work, where a similar analysis has been carried out, both in a nonrelativistic as well as in a relativistic setting, and with different geometries \cite{PLANO, HIPERBOLICO, NOSALTRESQUATRE, ESFERICO}. Thus the cosmological constant and the gravitational entropy of the matter contents of the Universe both qualify as quantum observables.

{}For completeness let us briefly summarise why quantum mechanics succeeds in encoding the key features of the cosmological fluid. For this purpose it is sufficiently accurate to regard spacetime as flat to begin with, although curvature corrections to the equations below are to be expected, which will be taken care of in a more thorough analysis \cite{FUTUR}. It is well known that the cosmological fluid, being essentially nonrelativistic, can be described as a Newtonian fluid governed by the continuity equation and the Euler equation:
\begin{equation}
\frac{\partial\rho}{\partial t}+\nabla\cdot\left(\rho{\bf v}\right)=0,\qquad \frac{\partial{\bf v}}{\partial t}+\left({\bf v}\cdot\nabla\right){\bf v}+\frac{1}{\rho}\nabla p-{\bf F}=0.
\label{oule}
\end{equation}
In the presence of viscosity, the Euler equation is replaced by the Navier--Stokes equation, but we will disregard this possibility. 

Less known is the fact that there is another nonrelativistic framework that can be mapped 1-to-1 into an ideal fluid dynamics, namely the Schroedinger quantum mechanics of a single scalar particle of mass $M$. Following Madelung \cite{MADELUNG} one factorises the wavefunction $\Psi$ into amplitude and phase: 
\begin{equation}
\Psi=\exp\left(S+\frac{{\rm i}}{\hbar}I\right).
\label{fatto}
\end{equation}
The phase $\exp({\rm i}I/\hbar)$ is the complex exponential of the classical--mechanical action integral $I$, and we have written the amplitude as the exponential of the real, dimensionless quantity $S$. The ideal fluid at work here is the {\it quantum probability fluid}\/. The latter is subject to the {\it quantum potential}\/ $Q$, which supplements the external potential $V$ present in the Schroedinger equation.  Altogether, this 1-to-1 correspondence between the quantum probability fluid and an Euler fluid is summarised in the following table:
\begin{equation}
\begin{tabular}{| c | c | c |}\hline
& Euler &  Madelung  \\ \hline
volume density& $\rho$ & $\exp(2S)$   \\ \hline
velocity & ${\bf v}$ & $\nabla I/M$   \\ \hline
pressure term & $\nabla p/\rho$ & $\nabla Q/M$   \\ \hline
external forces & ${\bf F}$ & $-\nabla V/M$   \\ \hline
\end{tabular}
\label{tavola}
\end{equation}
It follows that given a Newtonian cosmological fluid, one can use nonrelativistic quantum mechanics to describe it as a quantum probability fluid. In this work we will make use of this possibility in order to provide a quantum--mechanical analysis of the cosmological fluid within a positively--curved FLRW spacetime.\footnote{The quantity $2S$ in Eq. (\ref{tavola}) will turn out to be the dimensionless entropy ${\cal S}$ defined in Eq. (\ref{62}), {\it i.e.}\/, ${\cal S}/(2k_B)=S$, where $k_B$ is Boltzmann's constant. Note the different typographies for the dimensionful ${\cal S}$ and the dimensionless $S$.}

In our use of special functions we follow the conventions of ref. \cite{GRADSHTEYN}. Astrophysical data such as the numerical values of $R_U$ (radius of the observable Universe), $M$ (overall mass of the Universe), $\Lambda$ (cosmological constant) and $H_0$ (Hubble's constant) are taken from ref. \cite{PLANCK}.

\section{Metric, d'Alembertian and Laplacian}\label{due}

Our starting point is the spherical FLRW metric expressed in comoving coordinates $t$, $r$, $\theta$, $\phi$ \cite{WEINBERG2}:
\begin{equation}
{\rm d}s^2={\rm d}t^2-a^2(t)\,{\rm d}\Omega^2_3. 
\label{metrik}
\end{equation}
The term ${\rm d}\Omega^2_3$ is the standard metric on the 3--dimensional sphere $\mathbb{S}^3$ with radius $R_U$ (the radius of the observable Universe),\begin{equation}
{\rm d}\Omega^2_3=\frac{{\rm d}r^2}{1-\frac{r^2}{R_U^2}}+r^2\,{\rm d}\Omega^2_2,
\label{4}
\end{equation}
while
\begin{equation}
{\rm d}\Omega^2_2={\rm d}\theta^2+\sin^2\theta\,{\rm d}\phi^2
\label{3}
\end{equation}
is the round metric on the unit 2--dimensional sphere $\mathbb{S}^2$. The coordinate ranges covered are $0<r<R_U$, $0<\theta<\pi$, $0<\phi<2\pi$. 
The change of variables
\begin{equation}
r=R_U\sin\chi, \qquad 0<\chi<\pi
\label{2}
\end{equation}
conveniently reexpresses the metric (\ref{4}) as
\begin{equation}
{\rm d}\Omega^2_{3}=R_U^2\left({\rm d}\chi^2+\sin^2\chi\,{\rm d}\Omega^2_{2}\right).
\label{1}
\end{equation}
In these variables, all integrals over the sphere $\mathbb{S}^3$ carry the integration measure
\begin{equation}
\sqrt{g}\,{\rm d}^3x={R_U^3}\sin^2\chi\sin\theta\,{\rm d}\chi\,{\rm d}\theta\,{\rm d}\phi,
\label{52}
\end{equation}
and the d'Alembertian operator $\square_{\rm FLRW}$ with respect to the FLRW metric (\ref{metrik}) reads
\begin{equation}
\square_{\rm FLRW}=3\,\frac{\dot a(t)}{a(t)}\frac{\partial}{\partial t}+\frac{\partial^2}{\partial t^2}-\frac{1}{a^2(t)}\nabla_{\mathbb{S}^3}^2.
\label{5}
\end{equation}
Above, $\nabla^2_{\mathbb{S}^3}$ denotes the Laplacian on the 3--dimensional sphere $\mathbb{S}^3$:
\begin{equation}
\nabla_{\mathbb{S}^3}^2=\frac{1}{R_U^2\sin^2\chi}\left[\frac{\partial}{\partial\chi}\left(\sin^2\chi\frac{\partial}{\partial\chi}\right)+\frac{1}{\sin\theta}\frac{\partial}{\partial\theta}\left(\sin\theta\frac{\partial}{\partial\theta}\right)+\frac{1}{\sin^2\theta}\frac{\partial^2}{\partial\phi^2}\right].
\label{6}
\end{equation}

In the presence of an ideal fluid of density $\rho$ and pressure $p$, the Friedmann--Lema\^{\i}tre equations corresponding to the FLRW metric (\ref{metrik}) read
\begin{equation}
3\left(\frac{\dot a^2}{a^2}+\frac{1}{a^2}\right)=\kappa\rho+\Lambda,\qquad 2\,\frac{\ddot a}{a}+\frac{\dot a^2}{a^2}+\frac{1}{a^2}=-\kappa p+\Lambda,
\label{7}
\end{equation}
where $\kappa=8\pi G$ and a dot denotes ${\rm d}/{\rm d}t$. Given the extreme smallness of $\Lambda$ we may take 
\begin{equation}
3\left(\frac{\dot a^2}{a^2}+\frac{1}{a^2}\right)=\kappa\rho,\qquad 2\,\frac{\ddot a}{a}+\frac{\dot a^2}{a^2}+\frac{1}{a^2}=-\kappa p
\label{8}
\end{equation}
as a valid starting point; $\Lambda$ will be computed presently as a correction to Eq. (\ref{8}). Moreover it is usual to set $\dot a(t)/a(t)=H(t)$; as a first approximation we can assume $H(t)$ to be time--independent (Hubble's constant $H_0$). This yields the scale factor
\begin{equation}
a(t)={\rm e}^{H_0t}.
\label{9}
\end{equation}
The above scale factor is in excellent agreement with experimental data \cite{PERLMUTTER, RIESS} suggesting that the Universe currently finds itself in a period of exponential expansion. Using  (\ref{9}), the d'Alembertian $\square_{\rm FLRW}$ in Eq. (\ref{5}) becomes the d'Alembertian $\square_{H_0}$:
\begin{equation}
\square_{H_0}=3H_0\frac{\partial}{\partial t}+\frac{\partial^2}{\partial t^2}-{\rm e}^{-2H_0t}\nabla^2_{\mathbb{S}^3}.
\label{10}
\end{equation} 
One further consequence of the ansatz (\ref{9}), when substituted into Eq. (\ref{8}), is
\begin{equation}
\kappa\left(\rho+p\right)=2{\rm e}^{-2H_0t}.
\label{11}
\end{equation}
Asymptotically as $t\to\infty$ we recover $p=-\rho$, which is the equation of state for the cosmological constant \cite{WEINBERG2}. The extra factor of 2 on the right--hand side above is irrelevant for the late--time equation of state.

The eigenvalue problem corresponding to the d'Alembertian operator (\ref{10}),
\begin{equation}
\square_{H_0}\Psi_{\mu}(t,\chi,\theta,\phi)=\mu\Psi_{\mu}(t,\chi,\theta,\phi),
\label{40}
\end{equation}
is solved in the appendix. One finds the eigenfunctions and eigenvalues
\begin{equation}
\Psi_{\mu klm}(t,\chi,\theta,\phi)=T_{\mu k}(t)Y^{klm}(\chi,\theta,\phi),\qquad \mu\geq -9H_0^2/4,
\label{35}
\end{equation}
with explicit expressions for $Y^{klm}(\chi,\theta,\phi)$ and $T_{\mu k}(t)$ given in Eqs. (\ref{32}) and (\ref{31}), respectively.  The space--dependent piece $Y^{klm}(\chi,\theta,\phi)$ is square--integrable on the 3--dimensional sphere $\mathbb{S}^3$, while the constraint $\mu\geq -9H_0^2/4$ on the eigenvalues guarantees the square--integrability of $T_{\mu k}(t)$ over the half--line $t\in(0,\infty)$

It is important to observe that, after multiplication by $-\hbar^2/(2M)$, Eq. (\ref{40}) qualifies as a time--independent Schroedinger equation governing the cosmological fluid of a Universe modelled by the   quantum wavefunction $\Psi$. The corresponding time--dependent Schroedinger equation,
\begin{equation}
{\rm i}\hbar\frac{{\rm d}\Phi}{{\rm d}\tau}=-\frac{\hbar^2}{2M}\square_{H_0}\Phi,
\label{36}
\end{equation}
is satisfied by a $\tau$--dependent wavefunction $\Phi$ describing a free particle of mass $M$ given by
\begin{equation}
\Phi(\tau;t,\chi,\theta,\phi)={\rm e}^{-{\rm i}\mu\tau}\Psi_{\mu klm}(t,\chi,\theta,\phi).
\label{37}
\end{equation}
The Schroedinger time variable $\tau$ in Eqs. (\ref{36}), (\ref{37}) is not to be confused with the FLRW time variable $t$. As explained in ref. \cite{BEGGS2}, $\tau$ is best thought of as an external variable, independent of the cosmological time $t$. Thus the $\Psi_{\mu klm}(t,\chi,\theta,\phi)$ in (\ref{37}) are stationary states with respect to $\tau$, even if they depend explicitly on the FLRW time variable $t$. These $\Psi_{\mu klm}(t,\chi,\theta,\phi)$ are the quantum states that must be substituted into Eqs. (\ref{fatto}), (\ref{tavola}) for a quantum--mechanical picture of the cosmological fluid.

\section{The subspace of $t$--independent, radially symmetric quantum states}

In what follows we will restrict our attention to the $t$--independent piece $Y^{klm}(\chi,\theta,\phi)$ within the wavefunction $\Psi_{\mu klm}(t,\chi,\theta,\phi)$. The reason for this restriction is that we are interested in observables that are explicitly $t$--independent: then the $t$--dependent piece $T_{\mu k}(t)$ will drop out of any matrix element between any two states $\Psi_{\mu klm}(t,\chi,\theta,\phi)$.

Moreover,  the cosmological principle demands that we only consider spherically symmetric states, {\it i.e.}\/, states with $l=0$, $m=0$. Thus the subspace of $t$--independent, spherically symmetric quantum states is (the closure of) the linear span of the wavefunctions 
\begin{equation}
Y^{k00}(\chi,\theta,\phi)=\frac{\sqrt{2}}{2\pi}\frac{\sin(k+1)\chi}{\sin\chi}, \qquad k\in\mathbb{N},
\label{50}
\end{equation}
(see Eq. (\ref{miltres})). We will denote $Y^{k00}(\chi,\theta,\phi)$ more simply by $Y^k(\chi)$, since they are $\theta$, $\phi$ independent. Integration of the volume element (\ref{20}) over the unit, 2--dimensional sphere $\mathbb{S}^2$ yields the radial volume element $4\pi\sin^2(\chi){\rm d}\chi$ as the integration measure to be used within the subspace of radially symmetric wavefunctions. Thus the $Y^k(\chi)$ are orthonormal with respect to the effective inner product
\begin{equation}
\langle Y^k\vert Y^{k'}\rangle=4\pi\int_0^{\pi}{\rm d}\chi\,\sin^2\chi\left(Y^k(\chi)\right)^*Y^{k'}(\chi)=\delta^{kk'}, \qquad k,k'\in\mathbb{N}
\label{51}
\end{equation}
and complete within the subspace of radially symmetric wavefunctions. 

The product of $\vert Y^k(\chi)\vert^2$ times the measure $4\pi\sin^2\chi{\rm d}\chi$ equals the radial probability density $\rho_k(\chi)$ corresponding to the radially symmetric states (\ref{50}):
\begin{equation}
\rho_k(\chi)\,{\rm d}\chi=\frac{2}{\pi}\sin^2(k+1)\chi  \,{\rm d}\chi, \qquad k\in\mathbb{N}.
\label{19}
\end{equation}
The above density oscillates the least when $k=0$: this latter state is the radial vacuum, for which the minimal violation of the cosmological principle is attained. We will see below that the best fit to the measured value of the cosmological constant $\Lambda$ is reached by assuming the Universe finds itself in the quantum state $k=10$---not too far from the state that minimally violates the cosmological principle.

\section{Analysis of observables}\label{beable}

\subsection{The cosmological constant}\label{kosmokonst}

The operator 
\begin{equation}
\Lambda(r)=\frac{1}{r^2}=\frac{1}{R_U^2\sin^2\chi}
\label{landa}
\end{equation}
qualifies as a viable candidate  to represent the cosmological constant \cite{COSMONOI, MEX}. The matrix representing the above operator within the subspace of spherically symmetric states (\ref{50}) turns out to equal
\begin{equation}
\langle Y^{k}\vert\Lambda\vert Y^{k'}\rangle=\frac{2}{R_U^2}{\rm min}(k+1,k'+1)\delta(P_k,P_{k'}), \qquad k,k'\in\mathbb{N},
\label{60}
\end{equation}
where $\delta(P_k,P_{k'})$ equals $1$ (resp. $0$) if $k$ and $k'$ have the same (resp. opposite) parity. For an estimate of the orders of magnitude involved we set $k'=k$. Then
\begin{equation}
\langle Y^{k}\vert\Lambda\vert Y^{k}\rangle=\frac{2}{R_U^2}(k+1), \qquad k\in\mathbb{N}.
\label{61}
\end{equation}
A linear growth with the quantum number $k$ is in agreement with the results obtained in ref. \cite{ESFERICO} for a spherical but Newtonian spacetime. 

Applying Eq. (\ref{61}), the best fit to the current value of the cosmological constant, $\Lambda=1.1\times 10^{-52}$ metres$^{-2}$ or $\Lambda=2.9\times 10^{-122}$ in natural units, is attained for $k=10$ and a value of $R_U=4.4\times 10^{26}$ metres \cite{PLANCK}.

\subsection{Gravitational Boltzmann entropy}\label{gravbolent}

Let $M$  denote the total mass/energy contents of the Universe and $H_0$ the Hubble constant; we collect their numerical values from ref. \cite{PLANCK}:
\begin{equation}
M=3.0\times 10^{54}\,{\rm kg}, \qquad H_0=2.2\times 10^{-18}\, {\rm s}^{-1}.
\label{16}
\end{equation}
Specifically, $M$ is the sum of baryonic $M_{\rm bar}$, dark matter $M_{\rm dm}$ and dark energy $M_{\rm de}$ contributions: $M=M_{\rm bar}+M_{\rm dm}+M_{\rm de}$. We have the percentage fractions 
\begin{equation}
M_{\rm bar}=5M/100,\qquad M_{\rm dm}=27M/100,\qquad M_{\rm de}=68M/100.
\label{18}
\end{equation}
We apply the arguments put forward in ref. \cite{ESFERICO} and consider the operator
\begin{equation}
{\cal S}(r)=\frac{k_BMH_0}{\hbar}r^2=\frac{k_BMH_0}{\hbar}R_U^2\sin^2\chi
\label{62}
\end{equation}
as a measure of the gravitational Boltzmann entropy of the Universe. More precisely: with the cosmological fluid of the entire Universe described by a quantum wavefunction $\Psi$ in a curved FLRW background, the expectation value $\langle\Psi\vert {\cal S}\vert\Psi\rangle$ is the Boltzmann entropy of the matter fields in such a spacetime. In the basis (\ref{50}) of spherically symmetric states one finds\footnote{A quick guide to the computation: the factor $\sin^2\chi$ coming from $\sqrt{g}$ cancels the $\sin^{-2}\chi$ behaviour coming from $\vert Y^k\vert^2$, leaving a regular integrand.}  
\begin{equation}
\langle Y^k\vert\sin^2\chi\vert Y^{k'}\rangle=\frac{2}{\pi}\int_0^{\pi}{\rm d}\chi\,\sin^2\chi\sin(k+1)\chi\sin(k'+1)\chi.
\label{63}
\end{equation}
Although the above integral can be evaluated explicitly, again it is enough to  set $k'=k$ in order to estimate the orders of magnitude involved. When $k>0$ we arrive at
\begin{equation}
\langle Y^k\vert {\cal S}\vert Y^{k}\rangle =\frac{k_BMH_0R_U^2}{2\hbar}
\label{64}
\end{equation}
regardless of the radial quantum number $k>0$. Substituting the numerical values of $M$, $H_0$ and $R_U$ we find
\begin{equation}
\langle Y^k\vert {\cal S}\vert Y^{k}\rangle\sim 10^{101}\ll 10^{123},\qquad k>0,
\label{65}
\end{equation}
in comfortable compliance with the holographic principle \cite{THOOFT, SUSSKIND}.\footnote{When $k=0$, the right--hand side of Eq. (\ref{64}) gets corrected by a factor of $3/2$, but the holographic bound (\ref{65}) continues to be satisfied.}

\section{Discussion}

General relativistic quantum mechanics, cosmological quantum mechanics and the quantum Friedmann equations have been discussed recently in refs. \cite{BEGGS1, BEGGS2, MATONE3, MATONE2}. These approaches to quantum theory share some common features with our present work that we elucidate next. 

A free Hamiltonian operator is proportional to the Laplacian on the space in which the particle propagates. As opposed to Newtonian theories, where the time variable $t$ is a universal parameter, in relativistic theories $t$ qualifies as a coordinate. As such it enters the Laplacian, hence also the Hamiltonian for a free particle evolving in FLRW spacetime. The Laplacian used here is given in Eq. (\ref{10}); however, the time variable $t$ cannot be that of the Schroedinger equation (\ref{36}) governing the time evolution of quantum states. Instead, as suggested in ref. \cite{BEGGS2}, one needs an {\it external time}\/, here denoted by $\tau$, according to which quantum states undergo Schroedinger evolution. To reiterate: it is essential to distinguish between\\
{\it i)} $t$ as a coordinate entering the background geometry and operators;\\
{\it ii)} $\tau$ as an external parameter governing Schroedinger evolution.\\
In our previous analysis of quantum--mechanical observables on FLRW spacetime we considered the flat case \cite{NOSALTRESQUATRE}. In that work we did not distinguish between $t$ and $\tau$, but the conclusions of that paper remain valid all the same. The reason is that the operators considered in ref. \cite{NOSALTRESQUATRE} were all $t$--independent. Hence, when computing matrix elements and expectation values in \cite{NOSALTRESQUATRE}, the $t$--dependent piece of the wavefunction dropped out. The same phenomenon happens in positive curvature as treated in this work: the $t$--dependent piece of the wavefunction (Eqs. (\ref{31}), (\ref{33})) can be ignored. This is so because our two basic observables, the cosmological constant and Boltzmann entropy, are both $t$--independent. 

Of course, observables ${\cal O}(t)$ depending explicitly on $t$ will necessarily involve the functions $T_{\mu k}(t)$ of Eq. (\ref{31}). This will be the case when dealing with physical quantities assumed to vary with cosmological time $t$. While the standard $\Lambda$--CDM model assumes a constant $\Lambda$, recent observational hints \cite{YIN} suggest that dark energy might evolve, potentially meaning that $\Lambda$ is not truly constant but varies with time or redshift, possibly linked to scalar fields or phase transitions. Since this is not the primary purpose of the present paper, we defer such an analysis for a future publication \cite{FUTUR}.

Restriction to the subspace $l=0$, $m=0$ (radial subspace) within Hilbert space is imposed on us by the assumption of isotropy underlying the cosmological principle. This is a natural restriction to consider when dealing with radial observables such as those studied here. As was the case with the time variable $t$, any $\theta,\phi$ dependent piece within the wavefunction drops out of the computation of matrix elements of radially symmetric operators. While angular dependence cannot play a role under the assumption of isotropy, one may of course consider observables ${\cal O(\theta,\phi)}$, the matrix elements of which will necessarily involve the spherical harmonics $Y^{lm}(\theta,\phi)$. The multipole expansion of the CMB, not analysed here, is probably the most prominent example.

The physical interpretation of $k$ is that of a {\it radial quantum number}\/. As such it is analogous to the quantum number $n$ in the hydrogen atom. To lowest order of approximation, the energy eigenvalues of the hydrogen atom are independent of the angular quantum numbers $l,m$ and depend only on $n$. Analogously to the case of the hydrogen atom, our eigenvalue spectrum is purely discrete and depends only on $k$. Correspondingly, the quantum number $k$ is preferred because we have limited our analysis to radially symmetric states. 

Concerning the physical status of the choice $k=10$ made for the quantum number $k$ in Eq. (\ref{61}), elementary analysis shows that this value provides the best fit to the actual measured value of the cosmological constant. As such, the choice $k=10$ is intended primarily as an illustrative state showing that the observed $\Lambda$ can be realised within our framework. In connection with this, and regarding Eq. (\ref{65}), we should stress that satisfaction of the holographic bound is a property of the background parameters $R_U,H_0,M$ rather than being specific to the state $k=10$. In particular, the expectation value of the entropy (\ref{64}) is $k$--independent, and the holographic bound is respected by the entire Hilbert space of excited states. This implies that satisfaction of the bound cannot be used as a selection criterion for the quantum number $k$. Whether or not a dynamical or statistical argument could single out the value $k=10$, or some narrow band of values around it, remains unclear at the moment, and will be the subject of future analysis \cite{FUTUR}. 

We finally observe that our Eq. (\ref{40}) reduces to the quantum Friedmann equation presented in ref. \cite{MATONE2} upon setting $\mu=0$ and $H_0=0$ in our (\ref{40}). This is an intriguing observation that certainly deserves further analysis \cite{FUTUR}.

\section{Appendix}

Here we provide some computational details omitted in the body of the text.

The eigenvalue problem corresponding to the d'Alembertian (\ref{10}),
\begin{equation}
\square_{H_0}\Psi_{\mu}(t,\chi,\theta,\phi)=\mu\Psi_{\mu}(t,\chi,\theta,\phi),
\label{12}
\end{equation}
is solved as usual by separation of variables: 
\begin{equation}
\Psi(t,\chi,\theta,\phi)=T(t)\psi(\chi,\theta,\phi).
\label{14}
\end{equation}
Substituting (\ref{14}) into (\ref{12}) and denoting the separation constant by $C$ one arrives at the set of two equations
\begin{equation}
\nabla^2_{\mathbb{S}^3}\psi=C\,\psi
\label{22}
\end{equation}
and
\begin{equation}
\frac{{\rm d}^2T}{{\rm d}t^2}+3H_0\frac{{\rm d}T}{{\rm d}t}-\left(C\,{\rm e}^{-2H_0t}+\mu\right)\,T=0.
\label{25}
\end{equation}

\subsection{Space--dependent part $\psi$ of the eigenfunctions $\Psi$}

We will first solve Eq. (\ref{22}). Demanding that $\psi$ be nonsingular at $\chi=0$ and $\chi=\pi$ ({\it i.e.}\/, at $r=0$), the spatial wavefunction $\psi(\chi,\theta,\phi)$ will belong to the Hilbert space $L^2\left(\mathbb{S}^3, {\rm d}\mu(\chi,\theta, \phi)\right)$. Here the integration measure 
\begin{equation}
{\rm d}\mu(\chi,\theta, \phi)=\frac{1}{R_U^3}\sqrt{g}\,{\rm d}^3x=\sin^2\chi\sin\theta\,{\rm d}\chi\,{\rm d}\theta\,{\rm d}\phi
\label{20}
\end{equation}
is that dictated by the metric (\ref{1}) on the {\it unit}\/ sphere $\mathbb{S}^3$ ({\it i.e.}\/, after setting $R_U=1$), and the scalar product is given by
\begin{equation}
\langle\psi_1\vert\psi_2\rangle=\int_0^{\pi}{\rm d}\chi\,\int_0^{\pi}{\rm d}\theta\,\int_0^{2\pi}{\rm d}\phi\,\sin^2\chi\sin\theta\,\psi_1^*\psi_2.
\label{13}
\end{equation}
The spectrum of allowed values for the separation constant $C$ in (\ref{22}) turns out to be \cite{AVERY}
\begin{equation}
C=-\frac{1}{R_U^2}k(k+2), \qquad k\in\mathbb{N},
\label{21}
\end{equation}
and the eigenfunctions  are given by hyperspherical harmonics $Y^{klm}(\chi,\theta,\phi)$, where
\begin{equation}
k\in\mathbb{N},\quad l\in\mathbb{N},\quad m\in\mathbb{Z}\qquad  {\rm  with}\qquad  l\leq k\quad {\rm and}\quad -l\leq m\leq l.
\label{41}
\end{equation}
Now $\mathbb{S}^3$ is the homogeneous manifold ${\rm SO}(4)/{\rm SO}(3)$, and the collection of all hyperspherical harmonics $Y^{klm}$ provides  basis vectors for a unitary representation of the Lie algebra ${\rm so}(4)$. The carrier space for this representation is $L^2\left(\mathbb{S}^3,{\rm d}\mu(\chi, \theta, \phi)\right)$. This representation is reducible, its irreducible components being the subspaces spanned by those states $Y^{klm}$ with fixed $k$, with $l\leq k$, and with $-l\leq m\leq l$. One readily verifies that these latter subspaces are $(k+1)^2$--dimensional.

Altogether, the Laplacian eigenvalue equation on the 3--dimensional sphere with radius $R_U$ reads
\begin{equation}
\nabla^2_{\mathbb{S}^3}Y^{klm}(\chi, \theta, \phi)=-\frac{1}{R_U^2}k(k+2)Y^{klm}(\chi, \theta, \phi).
\label{24}
\end{equation}

For ease of reference, below we collect some technical properties of hyperspherical harmonics  \cite{AVERY}. Endow the 3--dimensional unit sphere $\mathbb{S}^3$ with the metric (\ref{1}) (after setting $R_U=1$). Then the $Y^{klm}(\chi,\theta,\phi)$ are given explicitly by 
\begin{equation}
Y^{klm}(\chi, \theta, \phi)=N(k,l)\,\sin^l\chi\,C_{k-l}^{l+1}(\cos\chi)\,Y^{lm}(\theta,\phi),
\label{32}
\end{equation}
where  
\begin{equation}
N(k,l)=2^{l}l!\sqrt{\frac{2(k+1)(k-l)!}{\pi(k+l+1)!}}
\label{32a}
\end{equation}
is a normalisation factor, the $Y^{lm}(\theta,\phi)$ are standard spherical harmonics on the 2--dimensional unit sphere $\mathbb{S}^2$,
and the $C_n^{\alpha}(x)$ are Gegenbauer polynomials. The latter can be  expressed in terms of the Gauss hypergeometric function $F(a,b;c;x)$  \cite{GRADSHTEYN}:
\begin{equation}
C_n^{\alpha}(x)=\frac{\Gamma(2\alpha+n)}{\Gamma(n+1)\Gamma(2\alpha)}\,F\left(-n,n+2\alpha;\frac{2\alpha+1}{2};\frac{1-x}{2}\right).
\label{graddy}
\end{equation}
In particular one has 
\begin{equation}
C_n^1(\cos\chi)=\frac{\sin(n+1)\chi}{\sin \chi},
\label{mildos}
\end{equation}
hence
\begin{equation}
Y^{k00}(\chi,\theta,\phi)=\frac{\sqrt{2}}{2\pi}\frac{\sin(k+1)\chi}{\sin\chi}, \qquad k\in\mathbb{N}
\label{miltres}
\end{equation}
which only depends on $\chi$.

The set of all hyperspherical harmonics is orthonormal
\begin{equation}
\int_{\mathbb{S}^3}{\rm d}\mu(\chi,\theta, \phi)\,\left[Y^{klm}(\chi, \theta,\phi)\right]^*Y^{k'l'm'}(\chi, \theta,\phi)=\delta^{kk'}\delta^{ll'}\delta^{mm'}
\label{mil}
\end{equation}
and complete
\begin{equation}
\sum_{k=0}^{\infty}\sum_{l=0}^{k}\sum_{m=-l}^l\left[Y^{klm}(\chi, \theta,\phi)\right]^*Y^{klm}(\chi', \theta',\phi')\label{miluno}
\end{equation}
$$
=\frac{\delta(\chi-\chi')\delta(\theta-\theta')\delta(\phi-\phi')}{\sin^2\chi\sin\theta}
$$
within the Hilbert space $L^2\left(\mathbb{S}^3,{\rm d}\mu(\chi, \theta, \phi)\right)$.

We summarise the previous information in the following chart:

\begin{equation}
\begin{tabular}{| c | c |}\hline
manifold &  $\mathbb{S}^3$  \\ \hline
coordinates &  $0<\chi<\pi$, $0<\theta<\pi$, $0<\phi<2\pi$  \\ \hline
metric &  ${\rm d}s^2={\rm d}\chi^2+\sin^2\chi ({\rm d}\theta^2+\sin^2\theta\,{\rm d}\phi^2)$  \\ \hline
Laplacian eigenvalue equation & $\nabla^2f(\chi, \theta, \phi)=C f(\chi, \theta, \phi)$ \\ \hline
Laplacian eigenfunctions &  $f(\chi, \theta, \phi)=Y^{klm}(\chi, \theta, \phi)=\langle \chi, \theta, \phi\vert klm\rangle$  \\ \hline
Laplacian eigenvalues &  $C=-k(k+2)$  \\ \hline
quantum numbers  & $k\in\mathbb{N}\quad l\in\mathbb{N}$, $l\leq k\quad m\in\mathbb{Z}$,  $-l\leq m\leq l$  \\ \hline
integration measure & ${\rm d}\mu=\sin^2\chi\sin\theta\,{\rm d}\chi\,{\rm d}\theta\,{\rm d}\phi$   \\ \hline
Hilbert space & $L^2(\mathbb{S}^3,{\rm d}\mu)$   \\ \hline
orthonormality  & $\langle klm \vert k'l'm'\rangle=\delta^{kk'}\delta^{ll'}\delta^{mm'}$   \\ \hline
completeness  & $\sum_{klm}\vert kml\rangle\langle klm\vert={\bf 1}$   \\ \hline
\end{tabular}
\label{tabulauno}
\end{equation}

\subsection{Time--dependent part $T$ of the eigenfunctions $\Psi$}

{}Finally we turn our attention to the time--dependent piece of $\Psi$ in (\ref{14}). Substitution of the eigenvalues (\ref{21}) into Eq. (\ref{25}) yields
\begin{equation}
\frac{{\rm d}^2T}{{\rm d}t^2}+3H_0\frac{{\rm d}T}{{\rm d}t}+\left[\frac{k(k+2)}{R_U^2}\,{\rm e}^{-2H_0t}-\mu\right]\,T(t)=0.
\label{trece}
\end{equation}
Two linearly independent solutions of the above are
\begin{equation}
T_{\mu k}^{(\pm)}(t)={\rm e}^{-\frac{3}{2}H_0t}\,J_{\pm\frac{1}{2H_0}\sqrt{9H_0^2+4\mu}}\left[\frac{1}{H_0R_U}\sqrt{k(k+2)}\,{\rm e}^{-H_0t}\right],
\label{30}
\end{equation}
where $J_p(z)$ is a Bessel function of the first kind. For regularity as $t\to\infty$ we pick the positive sign in (\ref{30}) and denote the remaining $T_{\mu k}^{(+)}(t)$ more simply by $T_{\mu k}(t)$:
\begin{equation}
T_{\mu k}(t)=N_{\mu k}^{-1/2}\,{\rm e}^{-\frac{3}{2}H_0t}\,J_{\frac{1}{2H_0}\sqrt{9H_0^2+4\mu}}\left[\frac{1}{H_0R_U}\sqrt{k(k+2)}\,{\rm e}^{-H_0t}\right],
\label{31}
\end{equation}
where the normalisation factor $N_{\mu k}$ is such that
\begin{equation}
N_{\mu k}=\int_0^{\infty}{\rm d}t\,\vert T_{\mu k}(t)\vert^2. 
\label{33}
\end{equation}
The above normalisation integral converges whenever $\mu\geq -9H_0^2/4$; however its precise value is immaterial, as it drops out of all our computations.

\vskip0.5cm
\noindent
{\bf Acknowledgements}\\
The authors acknowledge support by project MCIN/AEI/10.13039/501100011033 and by ERDF, “A way of making Europe”, under project PID2024-162480OB-I00.

\end{document}